\documentclass[apj,onecolumn,numberedappendix]{emulateapj}
\usepackage{apjfonts}

\shorttitle{Dust in primeval galaxies} \shortauthors{Mao et al.}
\journalinfo{Accepted by ApJ.}

\begin{document}

\title{The Role of the Dust in Primeval Galaxies:\\
A Simple Physical Model for Lyman Break Galaxies and
Ly${\alpha}$ Emitters.}

\author{J. Mao$^{1}$, A. Lapi$^{1}$, G.L. Granato$^{2,1}$, G. de Zotti$^{2,1}$, and L. Danese$^{1}$}
\affil{$^{1}$ Astrophysics Sector, SISSA/ISAS, Via Beirut 2-4, I-34014 Trieste, Italy\\
$^{2}$ INAF, Osservatorio Astronomico di Padova, Vicolo dell'Osservatorio 5, I-35122 Padova, Italy.}

\begin{abstract}
We explore the onset of star formation in the early Universe,
exploiting the observations of high-redshift Lyman-break galaxies
(LBGs) and Ly$\alpha$ emitters (LAEs), in the framework of the
galaxy formation scenario elaborated by Granato et al. (2004)
already successfully tested against the wealth of data on later
evolutionary stages. Complementing the model with a simple,
physically plausible, recipe for the evolution of dust attenuation in
metal poor galaxies we reproduce the luminosity functions (LFs) of
LBGs and of LAEs at different redshifts. This recipe yields a much
faster increase with galactic age of attenuation in more massive
galaxies, endowed with higher star formation rates. These objects have therefore
shorter lifetimes in the LAE and LBG phases, and are more easily
detected in the dusty submillimeter bright (SMB) phase. The short
UV bright lifetimes of massive objects strongly mitigate the effect
of the fast increase of the massive halo density with decreasing
redshift, thus accounting for the weaker evolution of the LBG LF,
compared to that of the halo mass function, and the even weaker
evolution between $z\approx 6$ and $z\approx 3$ of the LAE LF. The
much lower fraction of LBGs hosting detectable nuclear activity,
compared to SMB galaxies, comes out naturally from the evolutionary
sequence yielded by the model, which features the coevolution of
galaxies and active nuclei. In this framework LAEs are on the average
expected to be younger, with lower stellar masses, more compact, and associated to
less massive halos than LBGs. Finally, we show that the intergalactic
medium can be completely reionized at redshift $z\approx 6-7$ by
massive stars shining in protogalactic spheroids with halo masses
from a few $10^{10}\, M_{\odot}$ to a few $10^{11}\, M_{\odot}$,
showing up as faint LBGs with magnitude in the range $-17\la
M_{1350}\la -20$, without resorting to any special stellar initial
mass function.
\end{abstract}


\section{Introduction}

The impressive recent advances in the observational studies of high
redshift galaxies and quasars have brought us up to the outskirts of
`dark ages', when the amount of neutral intergalactic hydrogen
starts becoming significant and tentative indications of the
presence of metal-free population III stars begin to show up. This
may give us some confidence that we are approaching the onset of the
galaxy formation history and provides key information on physical
processes driving the early galaxy evolution and on their influence
on the intergalactic medium (IGM).

Different techniques and selection criteria have been exploited to
find high-$z$ galaxies. A most efficient method is the Lyman break
dropout technique (Steidel \& Hamilton 1993), first applied to
select $z\approx 3$ galaxies (Steidel et al. 1996; Lowenthal et al.
1997) and subsequently extended to redshifts of up to $\approx 10$
(Steidel et al. 1999, 2003; Dickinson et al. 2004; Giavalisco et al.
2004; Ouchi et al. 2004; Bunker et al. 2004; Vanzella et al. 2005,
2006; Bouwens et al. 2005; Thompson et al. 2006; Bouwens \&
Illingworth 2006a,b).

Very deep narrow band imaging at selected redshift windows (Cowie \&
Hu 1998) has also been successfully used to search for Ly$\alpha$
emission up to $z\approx 7$ (Ouchi et al. 2003; Kashikawa et al. 2006b;
Shimasaku et al. 2006; Iye et al. 2006).

While these techniques are sensitive to unobscured or moderately
obscured objects, very dusty distant galaxies with extremely high
star formation rates (SFRs) have been detected by SCUBA $850\,\mu$m surveys
(see e.g. Chapman et al. 2005; Aretxaga et al. 2007). Substantial numbers of
passively evolving high-$z$ galaxies have also been found (Fontana
et al. 2006; Grazian et al. 2007).

The different selections emphasize different aspects of the early
galaxy evolution that need to be integrated in a unified view. In
this paper we present a simple physical model featuring an
evolutionary link between Ly$\alpha$ selected (LAE), Lyman break
(LBG), submillimeter bright (SMB), and passively evolving galaxies,
establishing quantitative relations between luminosities and basic
physical quantities such as stellar and halo masses, star formation
rates, galactic ages, metal abundances and dust attenuation\footnote{We follow
the convention, increasingly adopted in the recent literature, of using the
term `attenuation' rather than  `extinction' to indicate the difference in
magnitudes between the observed and intrinsic starlight, which is
the final result of the interplay between the dust properties and the complex
relative geometry of stars and dust. The term `extinction' is reserved to the
idealized situation in which a single source is attenuated by a thin screen of
dust.}. The model extends to earlier evolutionary phases the approach by Granato
et al. (2004), according to which the evolution of massive galaxies
is tightly linked to the growth of the supermassive BHs at
their centers. The model has successfully accounted for an extremely
large amount of observational data on galaxies and quasars (Granato
et al. 2004, 2006; Cirasuolo et al. 2005; Silva et al. 2005; Lapi et
al. 2006; see Appendix A for details).

An extension of the model to the earliest evolutionary phases, when
metals were first produced, requires a specific treatment of dust
attenuation, which plays an important role in the derivation of
high-$z$ galaxy properties, since even small amounts of dust
strongly affect the rest-frame UV emission. For example, a
correction for dust attenuation corresponding to $E(B-V) \approx 0.2$
would increase the estimate of the SFR at $z
\approx 6$ by a factor of about $10$. The presence of significant
attenuation in high-$z$ LBGs has indeed been reported by many authors
(e.g., Steidel et al. 1999; Shapley et al. 2001, 2006; Huang et al.
2005; Ando et al. 2006; Yan et al. 2006; Burgarella et al. 2006;
Eyles et al. 2007; Stark et al. 2007b), not surprisingly since the
dust formation in early galaxies is expected to occur on very short
timescales (Morgan \& Edmunds 2003).

The outline of the paper is the following. In \S~2 we present an
overview of the model and show that, with a simple recipe for the
evolution of dust attenuation during the earliest phases of the
galaxy lifetime, accounting for the luminosity vs. $E(B-V)$ correlation
found by Shapley et al. (2001), it nicely reproduces the
observational estimates of the LBG luminosity functions (LFs) at
different redshifts. In \S~3 we investigate the absorption effects
on the Lyman continuum emission, the evolution of the Ly$\alpha$
LF, and the contributions of young galaxies to the
cosmic re-ionization. Finally, in \S~4 we discuss our results and
summarize our conclusions.

Throughout the paper we adopt a flat cosmology with matter density
$\Omega_M =0.3$, Hubble constant $H_0=70$ km s$^{-1}$ Mpc$^{-1}$,
and normalization of the mass variance $\sigma_8=0.8$. Unless
otherwise specified, magnitudes are in the AB system.

\section{Overview of the model}

The rest-frame UV emission is proportional to the SFR
$\dot{M}_{\star}$. This is in turn determined by a
number of physical processes (such as cloud-cloud collisions,
cooling, fragmentation, energy inputs from quasars, supernovae [SNae] and
stellar winds, etc. etc.) which occur on different linear and time scales.
Even with the more sophisticated numerical techniques, one has to
resort to many simplifying assumptions to derive the star formation
history, since these processes are extremely complex and occur on
scales well below the currently achievable resolution of $N$-body$+$hydro
simulations (Iliev et al. 2006).

\begin{figure}
\epsscale{1.}\plotone{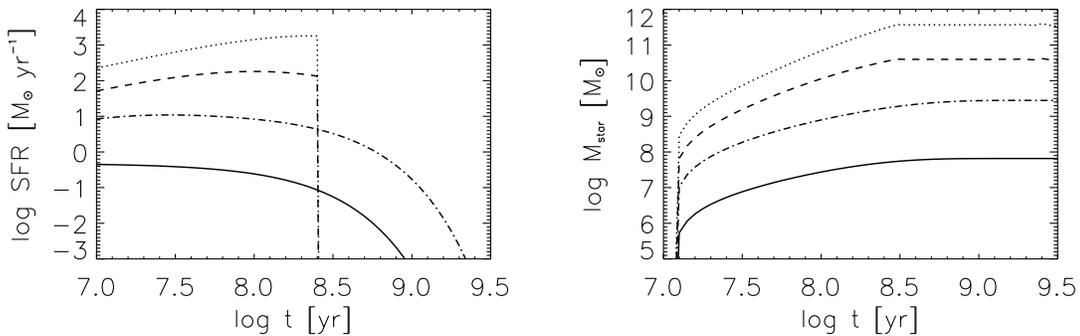}\caption{Left panel: SFR as a function
of the galaxy age for different halo masses $M_H=10^{10}\,
M_{\odot}$ (\textit{solid} line), $10^{11}\, M_{\odot}$
(\textit{dot-dashed} line), $10^{12}\, M_{\odot}$ (\textit{dashed}
line), and $10^{13}\, M_{\odot}$ (\textit{dotted} line), virialized
at $z=6$. Note the step-like cutoff in the SFR for the two higher masses,
due to the quasar feedback (see Appendix A for details).
Right panel: same for the stellar mass.}
\label{fig|sfr:mstar}
\end{figure}

In this paper we exploit the physical recipe elaborated by Granato
et al. (2004), which relates the average SFR to the distribution of
gas and dark matter (DM) in proto-galaxies, taking into account the
effects of the energy fed back to the intra-galactic gas by SN
explosions and by accretion onto the nuclear, supermassive black
hole (BH). The model envisages that during or soon after the
formation of the host DM halo, the baryons falling into the newly
created potential well are shock-heated to the virial temperature.
The hot gas is (moderately) clumpy and cools fast especially in the
denser central regions, where rapid mergers of a
fraction less than $30\%$ of the total mass yield strong bursts
of star formation that can reach up to thousands solar masses per yr
in massive halos at high redshift.
Such a huge star formation activity also promotes the storage of the
cooled gas into a reservoir around the central supermassive BH, eventually
leading to accretion onto it. The ensuing SN explosions and the nuclear activity
feed energy back to the gaseous baryons, and regulate the SFR and
the BH growth. These mutual energy feedbacks actually \emph{reverse} the
formation sequence of the stellar component of galaxies compared to
that of DM halos: the star formation and the buildup of central BHs
are completed more rapidly in the more massive haloes, thus
accounting for the phenomenon now commonly referred to as
\emph{downsizing}. The basic equations of the model and their
analytical solutions are reported in Appendix A.
The model predicts the time evolution of the SFR, of the
metal abundance $Z$ (taking into account the chemical enrichment of
the primordial infalling gas and the gas outflow due to energy
injection by SNae and active nuclei), of the stellar mass (see
Fig.~\ref{fig|sfr:mstar}), and of the mass stored in the central
supermassive BH.

When the star formation and the chemical evolution history of a
galactic halo of mass $M_H$ have been computed, the spectral energy
distribution (SED) as function of time from extreme UV to radio
frequencies is estimated through the GRASIL code (Silva et al.
1998). Coupling these results with the halo formation rate, we can
obtain the LFs of galaxies as a function of cosmic time. At $z\ga
2$, the halo formation rate can be reasonably well approximated by
the positive part of the derivative of the halo mass function with
respect to cosmic time (e.g., Haehnelt \& Rees 1993; Sasaki 1994).
The halo mass function derived from numerical simulations (Jenkins
et al. 2001) is well fit by the Sheth \& Tormen (1999, 2002)
formula, that improves over the original Press \& Schechter (1974)
expression, which is well known to under-predict by a large factor
the massive halo abundance at high redshift. Adopting the Sheth \&
Tormen (1999) mass function $N_{\mathrm{ST}}(M_H, T)$, $T$ denoting
the cosmic time, the formation rate of DM halos is given by
\begin{equation}\label{eq|DMrates}
{\mathrm{d}^2\, N_{\mathrm{ST}}\over \mathrm{d} M_H\, \mathrm{d}
T}=\left[{a\, \delta_c(T)\over \sigma^2(M_H)}+{2\, p\over
\delta_c(T)}\, {\sigma^{2\,p}(M_H)\over \sigma^{2\,p}(M_H)+ a^p\,
\delta_c^{2\,p}(T)}\right]\, \left|{\mathrm{d}\delta_c\over
\mathrm{d} T}\right|\, N_{\mathrm{ST}}(M_H, T)~,
\end{equation}
where $a=0.707$ and $p=0.3$ are constants obtained by comparison
with $N$-body simulations, $\sigma(M_H)$ is the mass variance of the
primordial perturbation field, computed from the Bardeen et al.
(1986) power spectrum with correction for baryons (Sugiyama 1995)
and normalized to $\sigma_8\approx 0.8$ on a scale of $8\,h^{-1}$
Mpc, and $\delta_c(t)$ is the critical threshold for collapse,
extrapolated from the linear perturbation theory.

Briefly, the model requires $10$ parameters (all of which are
constrained within rather small ranges by independent data or
physical arguments) to describe the physical processes ruling the
star formation, the BH growth and their feedback. With the parameter
values reported in Table 1 of Lapi et al. (2006), the model fits
the available data on the LFs of protospheroidal galaxies at
different redshifts and in different spectral bands, and the local
relationship between central BH mass and spheroidal luminosity
(hundreds of data points). Among others, the model reproduces the
850 $\mu$m extragalactic counts (Coppin et al. 2006; Scott et al.
2006) with their redshift distributions (Chapman et al. 2005;
Aretxaga et al. 2007), and correctly predicted (Silva et al. 2004,
2005) the fraction of high-$z$ galaxies in deep $24\,\mu$m surveys
(P\'erez-Gonz\'alez et al. 2005; Caputi et al. 2006; see
Magliocchetti et al. 2007), quantities that all proved to be very
challenging for competing models.

The evolving LFs of optical and X-ray selected quasars at  $z\ga
1.5$ are also well fitted by adding $3$ parameters, namely the
visibility times in the optical, $\Delta t_{\mathrm{vis}}=5\times
10^7\,$yr, and in X-rays, $\Delta t_{\mathrm{vis}}=3\times
10^8\,$yr, and a dispersion of $0.3$ dex around the mean
relationship between the halo and the central BH mass derived from
the model (see Lapi et al. 2006).

A summary of the model results and of their comparisons with
observational data is presented in Table 2 of Lapi et al. (2006).
To sum up, the model traces very well the formation and coevolution
of quasars and of their spheroidal hosts.

\begin{figure}
\epsscale{.7}\plotone{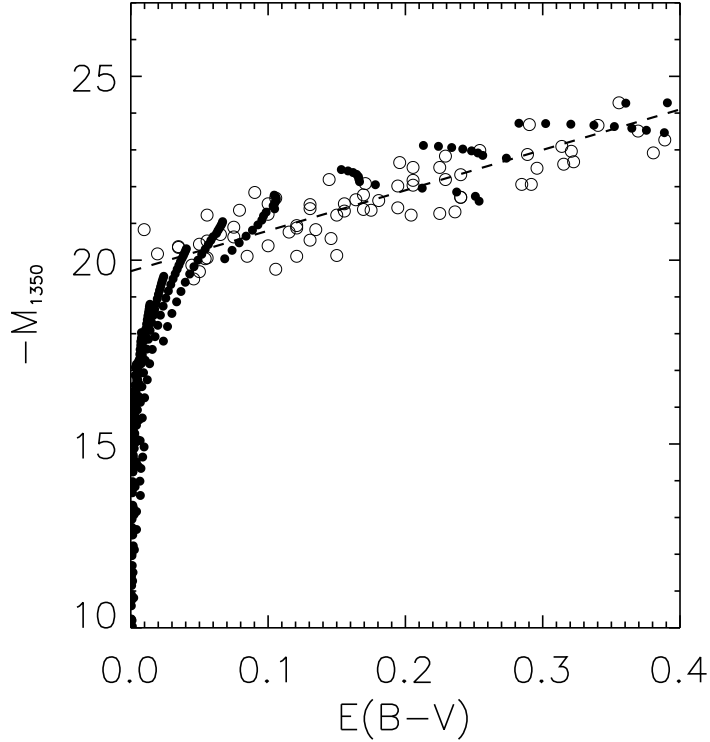}\caption{Correlation of the intrinsic
(attenuation-corrected) rest-frame absolute UV magnitude
$M_{1350}^{\mathrm{int}}$ with the color excess $E(B-V)$. The open
circles are the data by Shapley et al. (2001) at $z\approx 3$, and the dashed
line shows the average observational relationship. The black dots show
the expectations of our model using the attenuation
eq.~(\ref{eq|extgigi}), for halo masses in the range $10^{10}\,
M_{\odot}\la M_H \la 4\times 10^{12}\, M_{\odot}$ sampled in
intervals $\Delta \log M_H = 0.2$, and for ages in the range
$2\times 10^8\, \mathrm{yr}\la t \la \Delta t_{\mathrm{burst}}$ that correspond
to $\ga 70\%$ of the star-forming periods. The relation
$E(B-V)\approx A_{1350}/11$ by Calzetti et al. (2000) has been
adopted.}\label{fig|EBMV}
\end{figure}

\subsection{The UV luminosity function of LBG\lowercase{s}}

The absolute magnitude at $\lambda \approx
1350$ {\AA}, including attenuation, can be written as
\begin{equation}\label{eq|magn1350}
M_{1350}=51.59 - 2.5\, \log{\bar{L}_{1350}\over \mathrm{erg~s}^{-1}~
\mathrm{Hz}^{-1}} - 2.5\, \log{\dot{M}_{\star}\over M_{\odot}~
\mathrm{yr}^{-1}}+A_{1350}~,
\end{equation}
where $\bar{L}_{1350}\approx 1.2\times 10^{28}$ erg s$^{-1}$
Hz$^{-1}$ is the monochromatic luminosity for $\dot{M}_{\star} = 1
M_{\odot}~ \mathrm{yr}^{-1}$ (Kennicutt 1998, with the calibration
converted to the Romano et al. (2002) initial mass function [IMF]
adopted here, see Appendix A).

The dust attenuation is expected to be proportional both to some power
of star-forming gas mass (or equivalently of the SFR, see eq.~[\ref{eq|aSFR}])
that reflects the dust column density,
and to some power of the metallicity, that reflects the fraction
of metals locked into dust grains (see also Jonsson et al. 2006).
As shown by Fig.~\ref{fig|EBMV}, a simple, average relation between UV
attenuation, SFR and $Z$
\begin{equation}\label{eq|extgigi}
A_{1350}\approx 0.35\,\left(\frac {\dot{M}_{\star}}{M_{\odot}\,
\mathrm{yr}^{-1}}\right)^{0.45}\,
\left(\frac{Z}{Z_{\odot}}\right)^{0.8},
\end{equation}
consistent with expectations from the model, provides a very good
fit of the luminosity-reddening relation found by Shapley et al.
(2001). Variations of the normalization coefficient by $30\%$, of the
exponent of the SFR by $\pm 0.15$, and of the exponent of the metallicity by
$\pm 0.2$ are still compatible with the data.

\begin{figure}
\epsscale{1.}\plotone{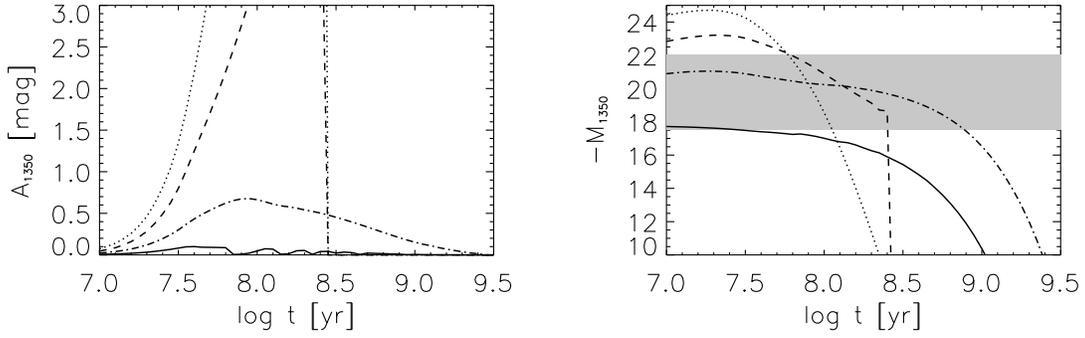}\caption{Attenuation (left panel) and
attenuated magnitude (right panel) at $1350$ {\AA} as a function of
galactic age, computed using eq.~(\ref{eq|extgigi})
and the quantities $\dot{M}_{\star}(t)$ and $Z(t)$ provided by our model.
The shaded area shows the range probed by the observed LFs. Lines are for different
halo masses virialized at $z=6$, as in
Fig.~\ref{fig|sfr:mstar}.}\label{fig|ext_magn}
\end{figure}

Using eq.~(\ref{eq|extgigi}) and the quantities $\dot{M}_{\star}(t)$ and
$Z(t)$ provided by our model (see Appendix A for handy approximations),
we compute the evolution with galactic age of the attenuation for LBGs with
different halo masses, see Fig.~\ref{fig|ext_magn} (left
panel). The attenuation is predicted to be small for LBGs hosted by
low-mass halos ($M_H\la 10^{11}\, M_{\odot}$), while it quickly
increases with time for higher masses, so that more massive galaxies
are UV-bright for shorter times. The more massive LBGs spend
approximately $90\%$ of their burst time in an interstellar medium
optically thick to their UV emission.  The right panel of
Fig.~\ref{fig|ext_magn} shows the predicted evolution of UV
luminosity with galactic age, including the attenuation given by
eq.~(\ref{eq|extgigi}), at $z=6$ for several halo masses.

The  UV LF of LBGs can be estimated coupling the UV luminosity and
halo mass derived from the our model (cf. Fig.~\ref{fig|ext_magn})
with the halo formation rate. The rest-frame UV LF at a cosmic time
$T$ is then
\begin{equation}\label{eq|UVcounts}
\Phi(M_{1350},T) = \int{\mathrm{d} M_{H}}~ {\mathrm{d}^2\,
N_{\mathrm{ST}}\over \mathrm{d} M_H\, \mathrm{d} T}\, {\mathrm{d}
\over \mathrm{d}M_{1350}}\, {\cal T}[M_{1350}|M_H,T]~,
\end{equation}
where ${\cal T}[M_{1350}|M_H,T]$ is the time spent at magnitudes
brighter than $M_{1350}$.

\begin{figure}
\epsscale{.7}\plotone{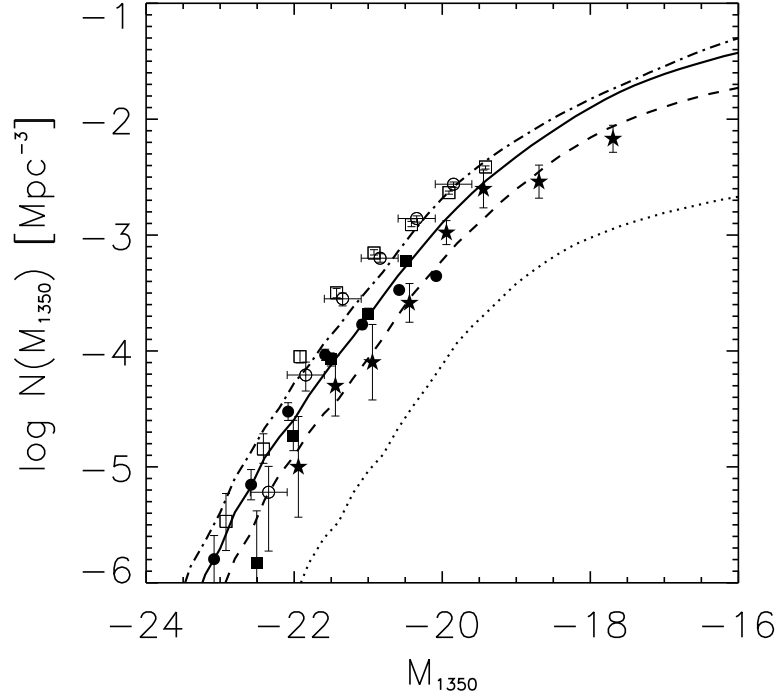}\caption{Model luminosity functions of
LBGs at $1350$ {\AA} for $z=3$ (\textit{dot-dashed} line), $z=5$
(\textit{solid} line), $z=7$ (\textit{dashed} line), and $z=10$
(\textit{dotted} line), based on the attenuation given by
eq.~(\ref{eq|extgigi}).  Data at $z\approx 3-4$ are from Steidel et
al. (2001; \textit{empty circles}) and Yoshida et al. (2006;
\textit{empty squares}); at $z\approx 4-5$ are from Iwata et al.
(2007; \textit{filled circles}) and Yoshida et al. (2006;
\textit{filled squares}); at $z\approx 5-7$ from Bouwens et al.
(2005; \textit{stars}). The Steidel data have been converted from
$1700$ {\AA} to $1350$ {\AA} setting $M_{1350}\approx M_{1700}+0.2$,
as suggested by Bouwens et al. (2005) on the basis of the composite
spectrum of $z\approx 3$ LBGs. }\label{fig|UVcounts_extgigi}
\end{figure}

As illustrated in Fig.~\ref{fig|UVcounts_extgigi}, good agreement
with the observational data is obtained. Neglecting attenuation would
increase the prediction at the bright end by more than an order of
magnitude. In fact dust attenuation is drastically shortening the
time ${\cal T}[M_{1350}|M_H,T]$ [eq.~(\ref{eq|UVcounts})] for large
halos (cf. Fig.~\ref{fig|ext_magn}). In this context it is  worth
noticing that the unattenuated LF can be depressed also  by assuming
short star formation timescales (around $80\,$Myr at $z\approx 6$),
as advocated by Stark et al. (2007a). On the other hand, such short
timescales would imply stellar masses and stellar ages much lower
than observed; Eyles et al. (2005, 2007) find stellar masses around
$1-3\times 10^{10}\, M_\odot$ and ages of hundreds Myrs. These
findings are quite naturally reproduced in our framework.

The approximate proportionality of the initial (galaxy age $t=0$)
SFR with the halo initial gas mass which, in turn, is proportional
to the halo mass [eq.~(\ref{eq|sfr})] implies that the highest
luminosity tail of the LF is associated to the most massive
galaxies. These are however very rare at high-$z$ and, furthermore,
their UV-bright phase is very short because of the quick increase of
their attenuation (Fig. \ref{fig|ext_magn}); they are thus easily
missed by the available surveys, covering very small areas. On the
other hand, low mass galaxies are too faint to be detectable. Hence,
according to the present model, the available data only sample the
halo mass range $10^{10}\, M_{\odot}\la M_H\la 10^{12}\, M_{\odot}$,
with SFRs of $\approx 1-100\, M_{\odot}\,\mathrm{yr}^{-1}$, and
stellar masses ranging from $\approx 10^7\, M_{\odot}$ to
$\mathrm{few}\times 10^{10}\, M_{\odot}$. The strong dependence of
$A_{1350}(t)$ on halo mass has clearly a key role in shaping the UV
LF.

The information on the LBG space densities at $z> 7$ is scanty.
Fig.~\ref{fig|UVcounts_extgigi} shows that we expect a drastic
decrease from $z = 7$ (dashed line) to $z = 10$ (dotted line).
Bouwens \& Illingworth (2006b) estimate that there are $1-4$ sources
brighter than $M_{1350}\approx -19.7$ at $7\la z \la 8$ on an area
of $5.8\,\mathrm{arcmin}^2$; we predict $2$ sources. Bouwens et al.
(2005) obtained an upper limit of $3$ galaxies at $z\approx 10$ on
an area of $15\,\mathrm{arcmin}^2$, to a magnitude limit of 28.5 in
$H_{160}$; our model yields $1$.

\section{The effect of dust on the Lyman-continuum of LBG\lowercase{s} and
on LAE\lowercase{s}}

\subsection{Escape fraction of ionizing photons}

The relative importance of dust attenuation and HI absorption in
determining the escape fraction of ionizing photons and the
Ly$\alpha$ luminosity depends on the complex geometry of the
distribution of gas, dust and stars. We assume that most of the
ionizing and of the Ly$\alpha$ photons originate in the very central
regions of the galaxy and see the same amount of dust as the photons
at the reference wavelength of $1350$ {\AA}; we extrapolate the
$1350$ {\AA} attenuation to shorter wavelengths adopting the
attenuation law proposed by Calzetti et al. (2000). This yields
$A_{1350}\approx 11\, E(B-V)$, $A_{1216}\approx 1.1\, A_{1350}$,
and, extrapolating from $1200$ {\AA} to $912$ {\AA} (cf. Draine
2003), $A_{912} \approx 1.6\,A_{1350}$.

The observed luminosity at $912$ {\AA} is given by
\begin{equation}\label{eq|L912}
L^{\mathrm{obs}}_{912}=L^{\mathrm{int}}_{912}\,f_{\mathrm{HI}}^{912}\,
f_{\mathrm{dust}}^{912}\, f_{\mathrm{IGM}}^{912}~,
\end{equation}
where $L^{\mathrm{int}}_{912}$ is the intrinsic luminosity at $912$ {\AA},
$f_{\mathrm{HI}}^{912}$ and $f_{\mathrm{dust}}^{912}\simeq e^{-A_{912}/1.08}$
are the fractions of $912$ {\AA} photons surviving $\mathrm{HI}$ and dust absorption
within the galaxy respectively, and
$f_{\mathrm{IGM}}^{912}$ is fraction surviving absorption
in the IGM along the line of sight.
The product $f_{\mathrm{dust}}^{912}\times
f_{\mathrm{HI}}^{912}\approx f_{\mathrm{esc}}$ is the fraction of $912$ {\AA} photons
emerging at the galaxy boundary.

Following Inoue et al. (2006) we define the ratio $R_{\mathrm{esc}}$
between the Lyman continuum and UV luminosity at the galaxy boundary as
\begin{equation}\label{eq|resc}
R_{\mathrm{esc}}=\frac{L_{912}^{\mathrm{obs}}\,
}{L_{1350}^{\mathrm{obs}}\, f_{\mathrm{IGM}}^{912}}~.
\end{equation}
The factor $f_{\mathrm{IGM}}^{1350}$, which in principle should appear in
this equation, is very close to unity and is therefore usually omitted.

Rest-frame ultraviolet spectroscopic observations in the Lyman
continuum region for samples of $29$ and $14$ LBGs at $z\approx 3$
have been obtained by Steidel et al. (2001) and by Shapley et al.
(2006). While Steidel et al. (2001) found $R_{\mathrm{esc}}\approx
0.22 \pm 0.05$, Shapley et al. (2006) found an average value about
$4.5$ times lower, $R_{\mathrm{esc}}\approx 0.05$, yet with a
substantial variance and with $2$ objects showing significant
emission below the Lyman limit, corresponding to
$R_{\mathrm{esc}}\approx 0.35$ and $0.22$. Note that these authors
determine $f_{\mathrm{IGM}}^{912}$ based on a sample of quasar spectra
at the same average redshift of the LBGs, and/or performing
numerical simulations of the line-of-sight absorption. A large
scatter of the observed ratio is expected, since the UV and the
ionizing radiation are emerging from central regions of galaxies
with quite complex gas and dust distributions. In fact, regions of
strong star formation in local galaxies exhibit line of sights with
quite variable dust attenuation and Ly$\alpha$ emission. In the
following we adopt $R_{\mathrm{esc}}\approx 0.15$ as a reference
value, but we explore the range $0.1\la R_{\mathrm{esc}}\la 0.2$.

\begin{figure}
\epsscale{.4}\plotone{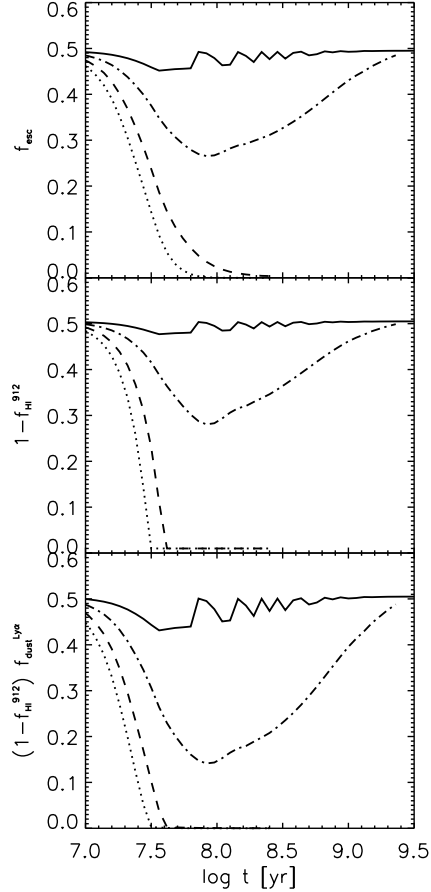}\caption{\textit{Top}: escape fraction
of ionizing photons as a function of galactic age. \textit{Middle}:
fraction of ionizing photons absorbed by HI within the galaxy as a
function of galaxy age. \textit{Bottom}: fraction of photons
escaping the galaxy at the Ly$\alpha$ wavelength. The lines
correspond to different halo masses virialized at $z=6$, as in
Fig.~\ref{fig|sfr:mstar}. In all panels we have adopted the
attenuation eq.~(\ref{eq|extgigi}), the Calzetti attenuation law
and $R_{\mathrm{esc}}=0.15$. The bumps in the solid lines are
just numerical artifacts.}\label{fig|fesc_fHI}
\end{figure}

Considering that $L^{\mathrm{obs}}_{1350}=L^{\mathrm{int}}_{1350}\,
f_{\mathrm{dust}}^{1350}$ (recall that $f_{\rm HI}^{1350}\approx f_{\rm IGM}^{1350}\approx 1$)
and setting $R_{\mathrm{int}}\equiv
L^{\mathrm{int}}_{912}/L^{\mathrm{int}}_{1350}$, the escape fraction
writes
\begin{equation}\label{eq|fesc}
f_{\mathrm{esc}}=\frac {R_{\mathrm{esc}}}{R_{\mathrm{int}}}\,
f_{\mathrm{dust}}^{1350}.
\end{equation}
The ratio between the intrinsic luminosities at $912$ {\AA} and at
$1350$ {\AA} for continuous star formation depends only on the
IMF and on the metal content. For the Romano
et al. (2002) IMF adopted here (see Appendix A for details), we find
$R_{\mathrm{int}}\approx 0.3$ (see also Leitherer et al. 1999; Steidel et al. 2001; Inoue et
al. 2005). In Fig.~\ref{fig|fesc_fHI} (top panel) we plot $f_{\mathrm{esc}}$
as a function of galactic age, computed according to the above equation and assuming
$f_{\mathrm{dust}}^{1350}\simeq e^{-A_{1350}/1.08}$;
it is apparent that for massive halos ($M_H \ga 10^{12}\,M_{\odot}$) the escape
fraction is negligible over most of the burst duration.

\begin{figure}
\epsscale{1.}\plottwo{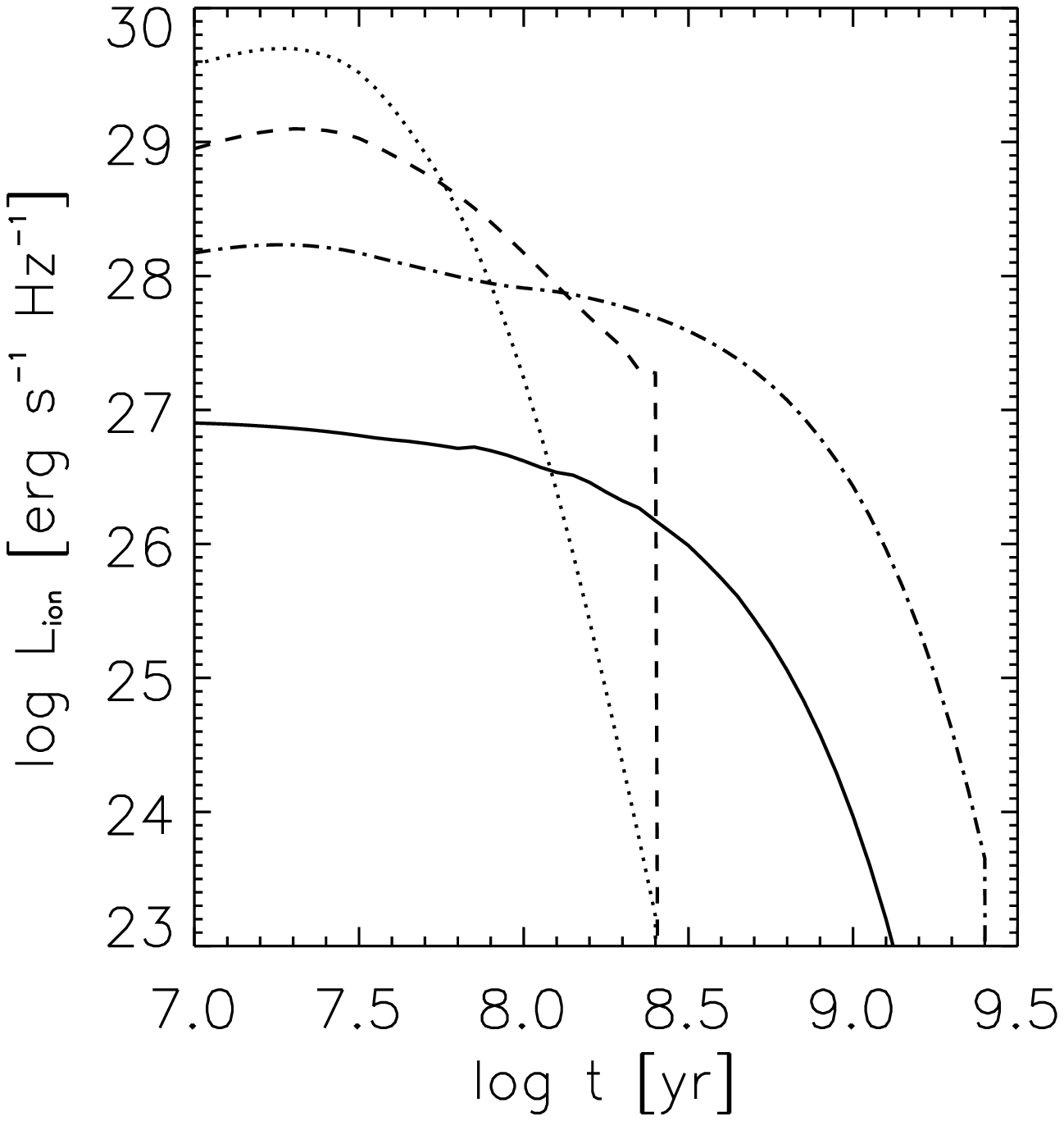}{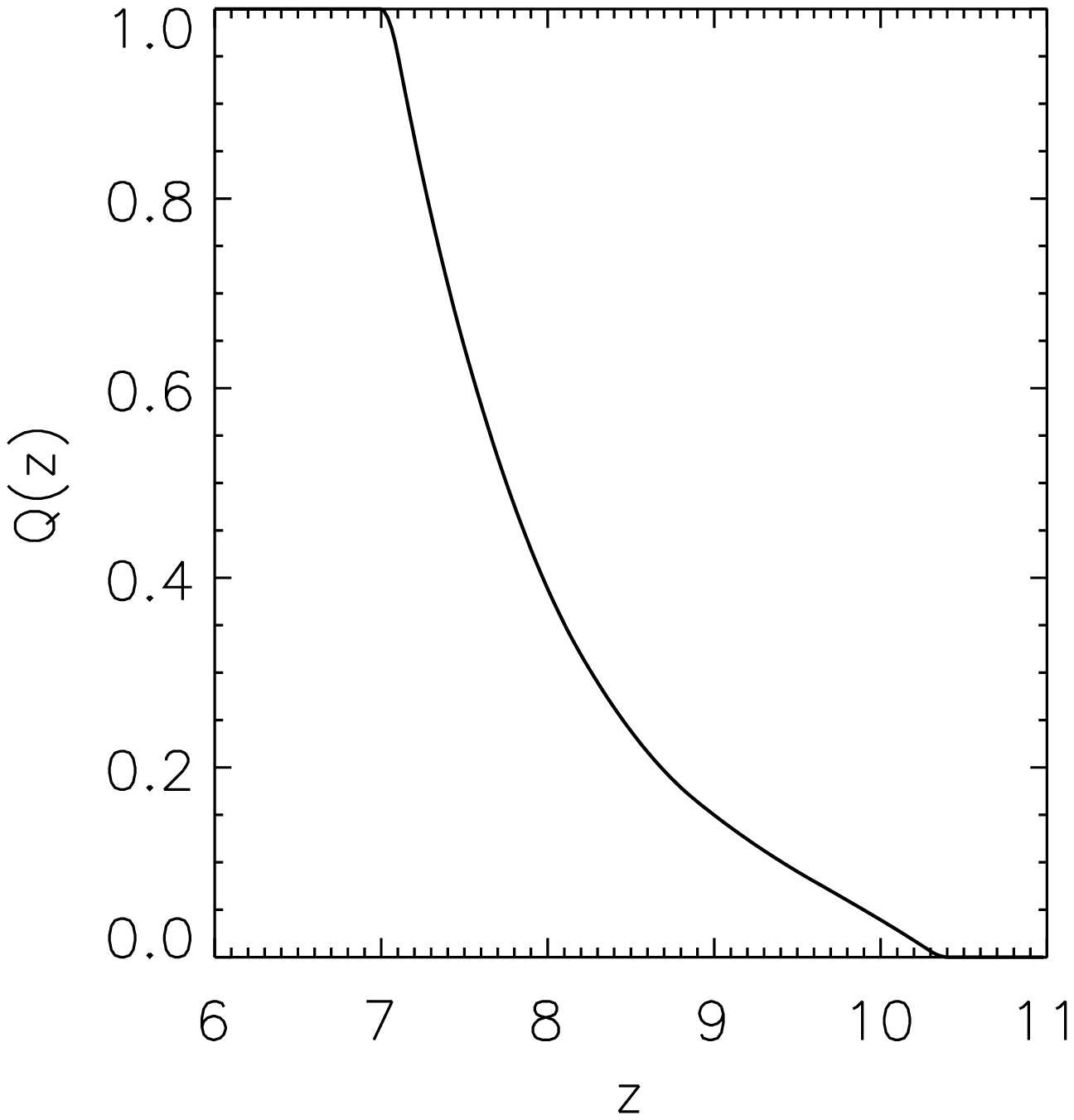}\caption{\textit{Left}:
evolution of the ionizing luminosity at $912$ {\AA} as a function of the galactic
age; lines correspond to different halo masses virialized at $z=6$,
as in Fig.~\ref{fig|sfr:mstar}. \textit{Right}: evolution of the
filling factor of ionized region in the Universe with the redshift.
In both panels we have adopted the attenuation
eq.~(\ref{eq|extgigi}), the Calzetti et al. (2000) attenuation law
and $R_{\mathrm{esc}}=0.15$.} \label{fig|reion_extgigi}
\end{figure}

On the other hand, the low mass (i.e. low intrinsic luminosity) LBGs
have large escape fractions and can thus release substantial amounts
of ionizing photons (Fig.~\ref{fig|reion_extgigi}, left panel).
The emission rate of ionizing $912$ {\AA} photons at the galaxy
boundary reads
\begin{equation}\label{eq|ndotion}
\dot{N}_{\mathrm{912}} = \frac{L_{912}}{h\nu _{912}}\approx
5.4\times 10^{53}\, f_{\mathrm{esc}}\,\left({\dot{M}_{\star}\over
M_{\odot}~\mathrm{yr}^{-1}}\right)~~\mathrm{s}^{-1}.
\end{equation}

The contribution of these galaxies to the cosmic reionization is
estimated in Appendix B. The filling factor of HII regions is
computed as a function of redshift (see
Fig.~\ref{fig|reion_extgigi}, right panel) for our reference
attenuation [eq.~(\ref{eq|extgigi})] and $R_{\mathrm{esc}}\approx 0.15$. It
turns out that LBGs are able to completely reionize the universe since
$z \approx 7$. The corresponding optical depth to
electron scattering is $\tau_{\mathrm{es}}\approx 0.07$, within the
range of the estimate from WMAP three year data
($\tau_{\mathrm{es}}=0.09\pm 0.03$, Page et al. 2007). Note that lowering
$R_{\rm esc}$ to $0.1$ and $0.05$ would move the completion of reionization to
$z\approx 6.6$ and $5.8$, respectively.

Most of the ionizing photons are produced by galaxies with halos
masses in the range $10^{10.6}\, M_{\odot} \la M_H \la 10^{11.4}\,
M_{\odot}$ at $z\la 8$. The contribution of massive galaxies is
depressed by their low number density at the relevant redshifts and
by their internal dust absorption. On the other hand, the
contribution of the numerous low mass galaxies is damped by the
decrease, driven by stellar feedback, of their SFRs. LBGs in the
above mass range have very low attenuation and contribute to the
faint part of the UV LF ($-20\la M_{1350}\la -18$) and to the bright
end of the observed Ly$\alpha$ LF at $z=6.5$ (see below).

\subsection{Ly$\alpha$ emitters}

The Ly$\alpha$ line typically arises from ionizing photons absorbed
by nearby hydrogen gas. If the fraction of absorbed ionizing photons
within the galaxy is ($1-f_{\mathrm{HI}}^{912}$) and $2/3$ of them are converted into
Ly$\alpha$ photons (Osterbrock 1989; see also Santos 2004), one gets
the unattenuated Ly$\alpha$ luminosity:
\begin{equation}\label{eq|Llyaint}
L^{\mathrm{int}}_{{\mathrm{Ly}}\alpha}= {2\over 3}\,
\dot{N}^{\mathrm{int}}_{912 }\,
(1-f_{\mathrm{HI}}^{912})\, h\nu_{{\mathrm{Ly}}\alpha}~.
\end{equation}
The interstellar dust attenuates
$L^{\mathrm{int}}_{{\mathrm{Ly}}\alpha}$ by a factor
$f_{\mathrm{dust}}^{\rm Ly\alpha}$;
furthermore, only a fraction
$f_{\mathrm{IGM}}^{\rm Ly\alpha}$ of Ly$\alpha$ photons survives the passage
through the IGM. Therefore the luminosity we see is
\begin{equation}\label{eq|Llyaobs}
L_{{\mathrm{Ly}}\alpha}^{\mathrm{obs}}\approx 5.6 \times 10^{42}\,
\left(\frac {\dot{M}_{\star}}
{M_{\odot}\,\mathrm{yr}^{-1}}\right)\, (1-f_{\mathrm{HI}}^{912}) \,
f_{\mathrm{dust}}^{\rm Ly\alpha}\, f_{\mathrm{IGM}}^{\rm Ly\alpha} ~~
\mathrm{erg~s}^{-1}~,
\end{equation}
for the adopted IMF; for a Salpeter IMF
the normalization should be reduced by a factor of about $1.6$.

Since the fraction of ionizing photons surviving
HI absorption is $f_{\mathrm{HI}}^{912}\approx f_{\mathrm{esc}}/f_{\mathrm{dust}}^{912}$,
eq.~(\ref{eq|fesc}) yields
\begin{equation}\label{eq|fHI}
f_{\mathrm{HI}}^{912}=\frac {R_{\mathrm{esc}}}{R_{\mathrm{int}}}\,
\frac{f_{\mathrm{dust}}^{1350}}{f_{\mathrm{dust}}^{912}}~.
\end{equation}
The fraction $f_{\mathrm{HI}}^{912}$ depends on the attenuation at
$1350$ {\AA} and on the attenuation law used to extrapolate it to
$912$ {\AA}. The middle and bottom panels of
Fig.~\ref{fig|fesc_fHI} show the absorbed fraction
($1-f_{\mathrm{HI}}^{912}$) and its product by
$f_{\mathrm{dust}}^{\rm Ly\alpha}$ as a function of the galactic age
for the attenuation eq.~(\ref{eq|extgigi}) and the Calzetti et
al. (2000) attenuation law; note that for the dust surviving fraction
we have assumed $f_{\mathrm{dust}}^{\lambda}\simeq e^{-A_{\lambda}/1.08}$.
It is apparent that LBGs in large
halos, $M_H\ga 10^{12}\, M_{\odot}$, are LAEs only for a short time
(less than $3 \times 10^7$ yr). The figure also illustrates the
difficulty of inferring the SFR from the Ly$\alpha$ luminosity: even
for relatively low mass halos ($M_H\la 10^{11}\, M_{\odot}$) the
effect of HI absorption and dust attenuation may be very large. This
implies that the Ly$\alpha$ emission from LBGs and in general the
statistics of LAEs at high redshifts are important
absorption/attenuation probes.

\begin{figure}
\epsscale{1.}\plottwo{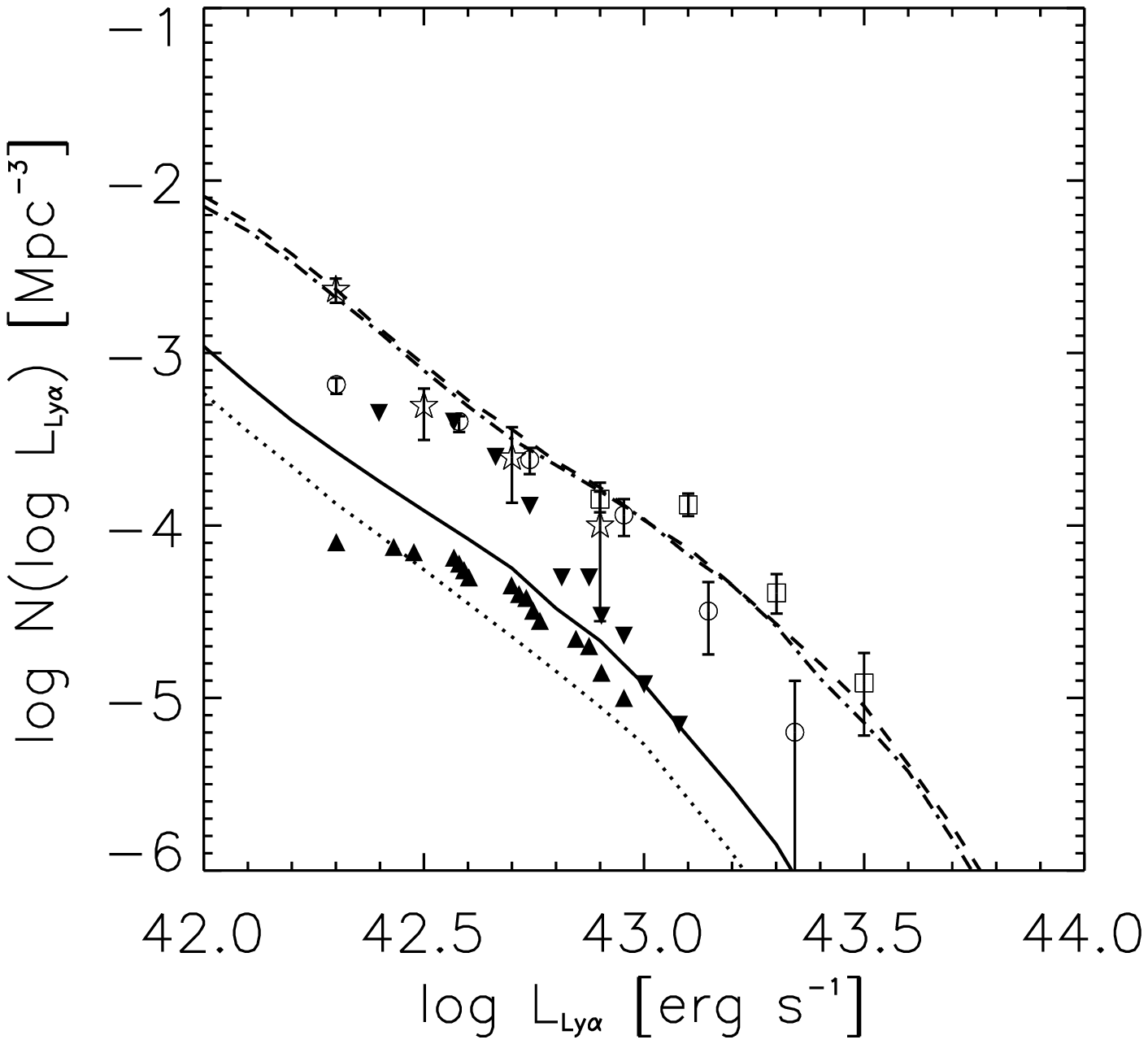}{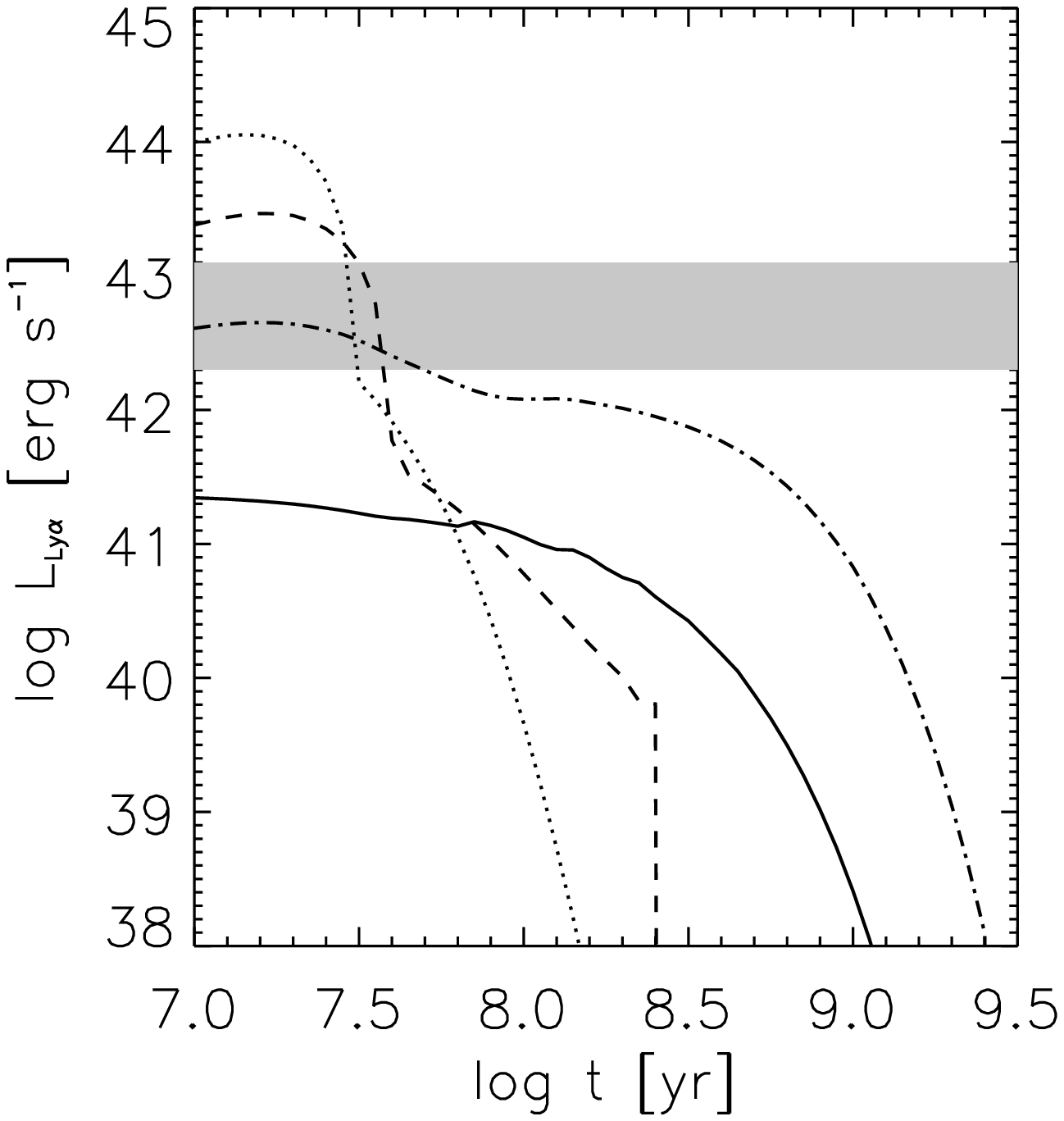}\caption{\textit{Left}: Model
Ly$\alpha$ LF for $z=4.9$ (\textit{dashed} line),
$z=5.7$ (\textit{dot-dashed} line), $z=6.4$ (\textit{solid} line),
and $z=8.2$ (\textit{dotted} line). Observational upper and lower
limits at redshift $z=6.4$ are from Kashikawa et al. (2006b;
\textit{filled triangles}), data at $z=5.7$ are from Shimasaku et
al. (2006; \textit{empty circles}) and from Murayama et al. (2007;
\textit{empty squares}), and data at $z=4.9$ are from Ouchi et al.
(2003; \textit{empty stars}). We have adopted the attenuation
eq.~(\ref{eq|extgigi}), the Calzetti et al. (2000) attenuation law,
$R_{\mathrm{esc}}=0.15$, $f_{\mathrm{IGM}}^{\rm Ly\alpha}=1$ at both $z=4.9$ and
$z=5.7$, and $f_{\mathrm{IGM}}^{\rm Ly\alpha}=0.5$ at both $z=6.4$ and $z=8.2$.
\textit{Right}: Ly$\alpha$ luminosity as a function of the galactic
age for $R_{\mathrm{esc}}=0.15$ and $f_{\mathrm{IGM}}^{\rm Ly\alpha}=0.5$. The
\textit{shaded area} shows the range probed by the observed LFs. The
lines correspond to different halo masses virialized at $z=6$, as in
Fig.~\ref{fig|sfr:mstar}.} \label{fig|Lya}
\end{figure}

Exploiting the relationship between the Ly$\alpha$ luminosity, the
SFR and the absorption [eq.~(\ref{eq|Llyaobs})], the LFs of LAEs at
high redshifts are estimated in the same way as the LBG LFs, and are
given by an equation strictly analogous to eq.~(\ref{eq|UVcounts}).
Our results are presented in Fig.~\ref{fig|Lya} (left panel).

It is clear from the above that the LF estimates rely on some
quantities that are not very well constrained observationally.
Still, it may be worth noticing that the data on the LF at $z\approx
4.9$ and $z\approx 5.7$ are consistent with a transmission factor
$f_{\mathrm{IGM}}^{\rm Ly\alpha}\approx 1$, for $R_{\mathrm{esc}} \la 0.15$, while
the LF at $z\approx 6.4$ seems to require $f_{\mathrm{IGM}}^{\rm Ly\alpha}\approx
0.5$. This may suggest that the ionization state of the IGM is
changing between $z\approx 5.7$ and $z \approx 6.4$ (see also
Kashikawa et al. 2006b). We need, however, to keep in mind that the
determination of $f_{\mathrm{IGM}}^{\rm Ly\alpha}$ is quite complex and implies
assumptions on the relative velocity of the emitting galaxies and
the IGM, on the presence of galactic winds, of neutral hydrogen,
etc. (see Haiman 2002, Santos 2004 and Dijkstra, Lidz \& Wyithe 2007
for comprehensive discussions). On the other hand, a fit of the
observed LF with $f_{\mathrm{HI}}^{912}\approx 0$ and ignoring dust
attenuation yields a conservative lower limit of $f_{\mathrm{IGM}}^{\rm Ly\alpha}\ga
0.1-0.4$ at $z\approx 6.5$ (see Dijkstra, Wyithe \& Haiman 2007).

Figure~\ref{fig|Lya} also shows our predictions for the LAE LF at
$z\approx 8$, when the Ly$\alpha$ line lies in the $J$ band (see
Barton et al. 2006), assuming $f_{\mathrm{IGM}}^{\rm Ly\alpha}\approx 0.5$.
However, if indeed the reionization is due to LBGs, we expect that
the IGM ionization level can be significantly lower at this
redshift, and the dimming of the Ly$\alpha$ luminosity substantially
stronger.

A clear-cut prediction of our model is that the Ly$\alpha$
luminosity of the most massive galaxies decreases abruptly after few
$10^7$ yr, implying that, with rare exceptions, bright LBGs should
exhibit relatively low Ly$\alpha$ luminosity (see
Fig.~\ref{fig|Lya},  right panel). A deficiency of objects with
large Ly$\alpha$ equivalent width among bright LBGs has recently
been reported by Ando et al. (2006).

According to the model, most of the LAEs with Ly$\alpha$
luminosities sampled by the available surveys ($2\times 10^{42}$ erg
s$^{-1}\la L_{{\mathrm{Ly}}\alpha}\la 3\times 10^{43}$ erg s$^{-1}$
at $z\ga 5-6$) are hosted by galactic halos within a rather narrow
mass range $5 \times 10^{10}\, M_{\odot} \la M_H \la 5 \times
10^{11}\, M_{\odot}$ and shining only for a short time $\la 8 \times
10^7$ yr. Fig.~\ref{fig|sfr:mstar} shows that the corresponding
stellar masses range from a few $10^{7}\, M_{\odot}$ to a few
$10^{9}\, M_{\odot}$, in nice agreement with the findings by
Filkenstein et al. (2007) and Pirzkal et al. (2007). Indeed, Gawiser
et al. (2006) noted that 80$\%$ of the objects in a sample of LAEs
at $z\approx 3.1$ have the right colors to be selected as LBGs. They
also inferred an average SFR around $6\, M_{\odot}$ yr$^{-1}$ and an
average mass in stars $M_{\star}\approx 5\times 10^8\,M_{\odot}$,
assuming a Salpeter IMF. Applying the correction by a factor of
about $2$ to account for the different IMFs, our model yields SFRs
and stellar masses consistent with these estimates. Gawiser et al.
(2006) also claim that the dust attenuation is on the average minimal
($A_V\la 0.1$) in their stacked sample. According to our model (cf.
Figs.~\ref{fig|sfr:mstar}, \ref{fig|ext_magn}, and \ref{fig|Lya}),
the attenuation is almost negligible for low stellar mass objects,
since both the SFR and the metal abundance are small. Only for
objects with $M_{\star}\ga 3 \times 10^8$ and age larger than
$5\times 10^7$ yr the attenuation becomes non negligible,
$A_{1350}\ga 0.3$. The dust in these objects may play an important
role in determining the EW of the Ly$\alpha$ (Filkenstein et al.
2007). However the effect is very sensitive to the spatial
distribution of dust, stars and HI.

Lai et al. (2007) found larger masses, from a few $\times 10^{9}\,
M_{\odot}$ to a few $\times 10^{10}\, M_{\odot}$, for $3$ LAEs at
$z\approx 6$. The corresponding ages vary from several Myr to
several hundreds Myr, depending on chemical composition. It has to
be noticed that these authors analyzed the 3 objects with the
largest fluxes at $3.6$ and $4.5$ $\mu$m. As a consequence we expect
that the selection is biased in favor of the objects with the
highest stellar mass. In our framework this selection corresponds to
objects with halo masses of $5-10 \times 10^{11}\, M_{\odot}$, ages
of about $40-50$ Myr (cf. Fig.~\ref{fig|sfr:mstar}), metallicities
around $1/3\, Z_{\odot}$ [cf. eq.~(\ref{eq|metallicity})] and
$E(B-V)\approx 0.15$ (cf. Fig.~\ref{fig|EBMV}); the predicted mass
in stars amounts to $4-5 \times 10^9\, M_{\odot}$.

In conclusion, the dust prescription we derive from observations of
LBGs is consistent also with the observations of LAEs at high
redshift.

\section{Discussion and conclusions}

We have extended to the earliest evolutionary phases of high
redshift galaxies the physical model worked out by Granato et al.
(2004) which proved to be capable of accounting for the wealth of
data on the heavily obscured active star formation phase probed by
(sub)mm surveys and on the subsequent passive evolution of
spheroidal galaxies. Lapi et al. (2006) have shown that this model,
which features a coevolution of galaxies and active nuclei, can
also provide good fits of the cosmic epoch dependent optical and
X-ray LFs of active galactic nuclei.

We have directly borrowed from the Granato et al. (2004) model three
basic ingredients: (i) the formation rate of galactic halos; (ii)
the star formation history as a function of halo mass and redshift,
and the IMF of stars; (iii) the chemical evolution. These
ingredients allow us to obtain the unattenuated
spectral energy distribution as a function of galactic age.
For the galaxies of interest here, which just begin to form their metals,
we need a specific recipe for the attenuation.
Based on simple physical arguments, we have adopted a power-law relation between
UV attenuation, SFR and metallicity [eq.~(\ref{eq|extgigi})], tuned to
fit the UV luminosity vs. $E(B-V)$ correlation found by Shapley et al.
(2001); this UV attenuation turns out to be increasing with galactic age and with
increasing galaxy mass. With this additional ingredient the model accounts also for
the observed LFs of LBGs at different redshifts.
The larger attenuation for more massive
objects implies that the UV LF evolves between $z\approx 6$ and $z\approx
3$ much less than the halo mass function. This is due to the fact
that the higher dust attenuation mitigates the fast increase with
decreasing redshift of the number density of massive halos. As a consequence,
the UV LF is only weakly sensitive to the values of cosmological
parameters, and in particular of $\sigma_8$. However, if LF
estimates will be extended to $z\approx 10$, the effect of
$\sigma_8$ could be visible down to small enough masses for which
the effect of attenuation is small.

The derived UV LFs imply that LBG can account for a complete reionization at $z \approx 7$. Most ionizing
photons are produced by galaxies with halo masses in the range
$10^{10.6}\, M_{\odot} \la M_H \la 10^{11.4}\, M_{\odot}$ at $z\la
8$, that dominate the faint portion of the UV LF.

We expect that LBGs with detectable Ly$\alpha$ emission are on average
much younger, less massive and less dusty, in agreement
with the analysis of $z\sim 4$ LBGs observed in the GOODS-S survey by Pentericci et al.
(2007). Note, however, that a second
Ly$\alpha$ bright phase may be expected at much later times, when
the interstellar medium is being swept away by super-winds driven by
the quasar feedback, allowing Ly$\alpha$ photons to escape from the
galaxy (see also Thommes \& Meisenheimer 2005).

\begin{figure}
\epsscale{1.}\plotone{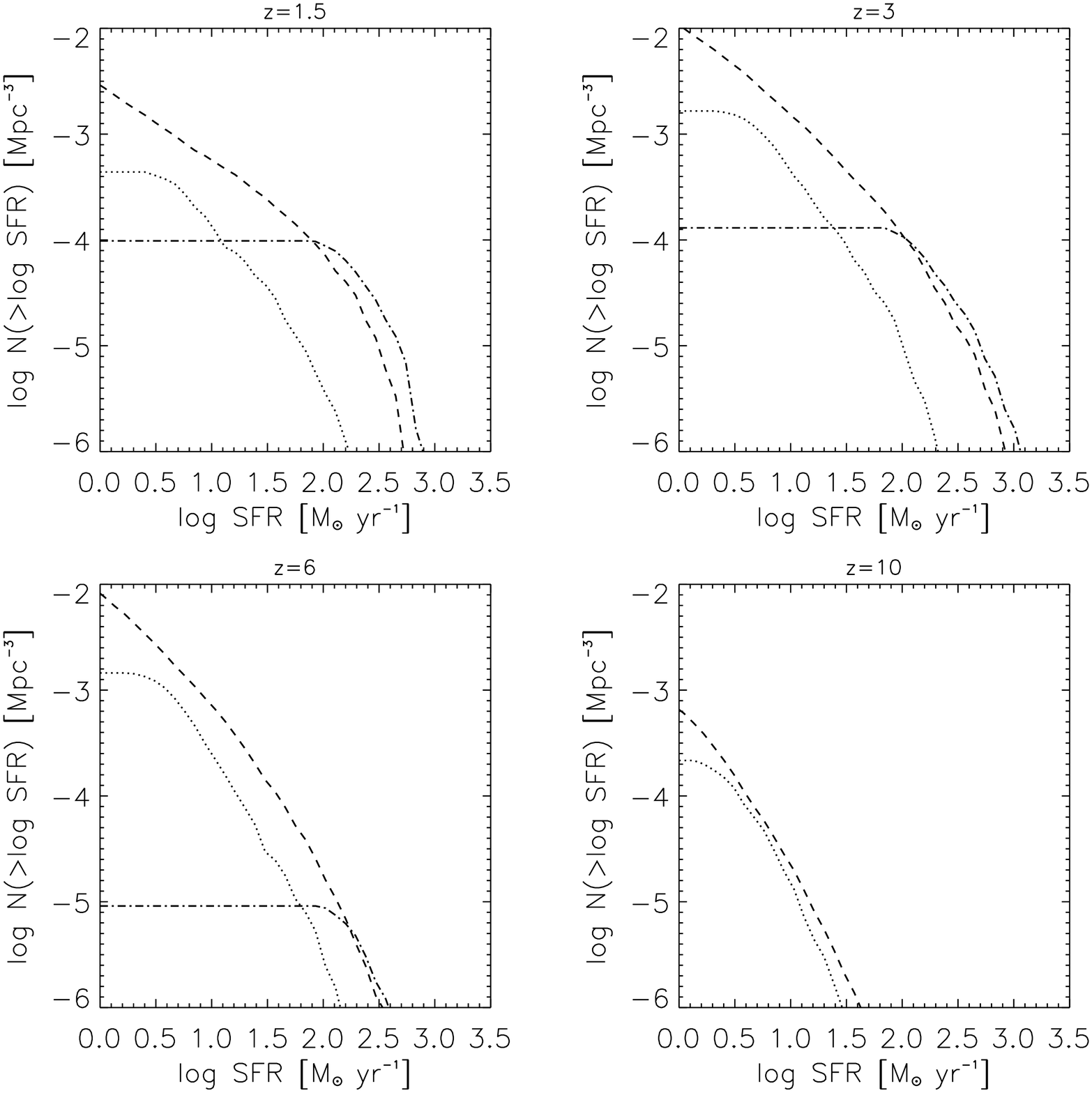}\caption{\textit{Left}: Cumulative
distribution of SFRs for LBGs with $-24\la M_{1350}\la -18$
(\textit{dashed} line), LAEs with $10^{42}\, \mathrm{erg~s}^{-1}\la
L_{\mathrm{Ly}\alpha}\la 3\times 10^{43}\, \mathrm{erg~s}^{-1}$
(\textit{dotted} line), and submm bright sources with $S_{850\,
\mu\mathrm{m}/(1+z)}\ga 1$ mJy (\textit{dot-dashed} line) at different
redshifts.}
\label{sfrcounts}
\end{figure}

Only a small fraction ($2-4\%$) of LBGs at $z\approx 3$ are detected
by deep X-ray surveys (Laird et al. 2006; Lehmer et al. 2005). X-ray
luminosities estimated using stacking techniques are generally below
$10^{42}\,\mathrm{erg}\,\mathrm{s}^{-1}$, consistent with the
luminosities due to the starburst activity (Type-II SN remnants, hot
gas, and high mass X-ray binaries).
The small fraction of $z\approx 3$ LBGs with $L_X \ga
10^{42}\,\mathrm{erg}\,\mathrm{s}^{-1}$ generally shows properties of active
nuclei. A single faint active nucleus was found by Douglas et al. (2007) in a large
sample of $z>5$ LBGs.
Our model implies that intense star formation
activity is accompanied by an exponential growth of a supermassive
BH at the galaxy center. The corresponding nuclear X-ray emission
(cf. Granato et al. 2006) can be written as:
\begin{equation}
L_X\approx 3\times 10^{44}\, \frac{\eta}{0.1}\,
\left(\frac{K_X}{20}\right)^{-1}\, \left({\dot{M}_{\bullet}\over
M_{\odot}~\mathrm{yr}^{-1}}\right)~~\mathrm{erg~s}^{-1},
\end{equation}
where $\eta$ is the mass-to-radiation conversion efficiency, $K_X$
is the bolometric correction for the $2-10$ keV band and
$\dot{M}_{\bullet}$ is the accretion rate onto the supermassive BH.
In the LBG phase the nuclear BHs are still very far from their final
masses (proportional to the halo masses) and only those associated
to halos at the upper limit of the range represented in the
available samples ($M_H\ga 10^{12}\, M_{\odot}$, see Fig.
\ref{fig|ext_magn}) can reach Eddington limited accretion rates
large enough ($\dot{M}_{\bullet}\ga 0.003\, M_{\odot}$ yr$^{-1}$) to
yield $L_X\ga 10^{42}$ erg s$^{-1}$ (see Fig. 1 of Lapi et al.
2006); the SFRs of these objects exceed $100\,M_{\odot}\, \mathrm{yr}^{-1}$ (Fig. \ref{fig|sfr:mstar}).
Interestingly, the model predicts that in more massive primeval galaxies
during the late SMB phase, the SFRs may reach thousands $M_{\odot}$ yr$^{-1}$
and correspondingly the nuclear accretion rates may attain tens $M_{\odot}$ yr$^{-1}$.
In fact, Granato et al. (2006) have shown that the model
successfully reproduces the X-ray observations of SMB galaxies
by Alexander et al. (2005a,b) and Borys et al. (2005).

We find that the model reproduces the Ly$\alpha$ LFs at $z=4.9$, $z=5.7$ and $z=6.5$ (Ouchi
et al. 2003; Shimasaku et al. 2006; Kashikawa et al. 2006b).
The effect of increasing opacity with increasing mass is even
stronger for LAEs and is responsible for the weaker evolution
of the LAE LF, compared to the LBG's, between $z=6$ and $z=3$,
clearly discernible in Fig.~\ref{fig|Lya}.
Evidences of only weak evolution of the LAE LF between $z=6.6$ and $z=3.4$
have been noted by Taniguchi et al. (2005; see also Murayama et al. 2007).

According to this model, the galaxies currently selected in Ly$\alpha$
($L_{\rm Ly\alpha}\ga 2\times 10^{42}$ erg s$^{-1}$)
are very young (cf. Fig.~\ref{fig|Lya}), with typical ages up to $\mathrm{few}\times 10^7\,$yr.
The Ly$\alpha$ luminosity range sampled by surveys carried out so far corresponds to a narrow
distribution of halo masses around $\simeq 10^{11}\, M_\odot$, and
stellar masses ranging from $5\times 10^7\, M_{\odot}$ to $10^9\, M_{\odot}$
(see Fig. \ref{fig|sfr:mstar}).
For given halo mass, Ly$\alpha$-selected galaxies
have lower stellar masses than LBGs and
their stars are preferentially formed in the densest central
regions; they are therefore expected to be more compact than LBGs,
as is indeed observed (Dow-Hygelund et al. 2007).
Galaxies with higher halo masses have higher
Ly$\alpha$ luminosities but are rare because of the low number
density of massive halos at high redshifts and of their short
lifetime in the LAE phase; they are thus easily missed by the
available surveys, covering very small areas. These rare objects
would however have very interesting properties: high Ly$\alpha$
luminosity with very young ages, low metallicities and
correspondingly high Ly$\alpha$ equivalent widths, low masses in
stars within large halo masses. On the other hand, the lower
luminosity portion of the LAE LF is dominated by relatively
older objects, more chemically evolved and with lower Ly$\alpha$
equivalent widths. Therefore the extension
to lower luminosities of the LAE LF will depend rather critically on
the minimum detectable equivalent width.
Both Ly$\alpha$-selected galaxies and LBGs are associated to relatively
small DM halos and so that they are not expected to show strong
clustering, consistent with the results by Murayama et al. (2007).

Fig.~\ref{sfrcounts} shows the cumulative distribution of SFRs for
LBGs, LAEs, and SMBs at different redshifts.
Although we must be aware that there is a significant
overlap among these populations, some trends clearly emerge. At high $z$ most of the
star formation is associated to the LBG phase, but a substantial
fraction of the galaxies with the highest star formation rates are
predicted to be dust-obscured (and therefore SMB).
This is because, in the present framework, the more massive galaxies have higher
SFRs, yielding higher metallicities and higher dust attenuations.
At redshift $z\ga 6$, however, the space density of SMBs
is very low. Only later on, an increasing fraction of the cosmic
SFR occurs in very dusty galaxies. Since SFRs are strongly correlated with halo
masses, SMB galaxies are expected to
show stronger clustering, consistent with the results by Blain et
al. (2004). Observations of the LBG angular correlation function at
$z\ga 4$ (Kashikawa et al. 2006a; Lee et al. 2006; Hildebrandt et
al. 2007) find evidences of substantial clustering, yet not as
strong as indicated by data on SMB galaxies. In particular, the clustering was
found to depend on luminosity, suggesting a close correlation
between halo mass and SFR, in keeping with our expectations.

To sum up, our model establishes a coherent scenario encompassing the
variety of galaxy populations selected by different techniques:
Ly$\alpha$ emitters, Lyman break galaxies, submillimeter bright, and
passively evolving galaxies. In this scenario, these populations
correspond to subsequent steps in the evolutionary sequence of
high-$z$ galaxies.

\begin{acknowledgements}
We warmly thank A. Bressan for enlightening discussions, and the
anonymous referee for very useful comments and suggestions. This work
is partially supported from ASI, INAF and MIUR grants.
\end{acknowledgements}

\begin{appendix}

\section{A simple recipe for the star formation rate in protogalaxies}

The history of star formation, of fueling of the mass reservoir
around the active nucleus, and of accretion into the central BH can
be easily computed by numerically solving the set of differential
equations written down by Granato et al. (2004; see also the
Appendix A of Lapi et al. 2006). In this Appendix we summarize the
results concerning the SFR, which are the relevant ones for this
paper, and, whenever possible, we give simple analytical
approximations for them.

At the virialization time a DM halo of mass $M_H$ hosting a galaxy
contains a mass $M_{\mathrm{inf}}(0)=f_{\mathrm{cosm}}M_{H}$ of
hot gas at the virial temperature, $f_{\mathrm{cosm}}\approx 0.18$
being the mean cosmic baryon to DM mass-density ratio. The gas
cools and flows toward the central regions of the halo at a rate
\begin{equation}\label{eq|mdotinf}
\dot{M}_{\mathrm{inf}}=
-\dot{M}_{\mathrm{cond}}-\dot{M}_{\mathrm{inf}}^{QSO}~~,
\end{equation}
where $\dot{M}_{\mathrm{cond}}=M_{\mathrm{inf}}/t_{\mathrm{cond}}$,
and the `condensation' timescale
$t_{\mathrm{cond}}=\max[t_{\mathrm{cool}}(R_H),t_{\mathrm{dyn}}(R_{H})]$
is the maximum between the dynamical time and the cooling time at
the halo virial radius $R_H$. The cooling time includes the
appropriate cooling function (Sutherland \& Dopita 1993) and allows
for a clumping factor ${C}$ in the baryonic component. The second
term on the right hand side describes the influence of the quasar
kinetic energy output on the hot gas distributed throughout the DM
halo. Note that in the above equation the effect of the angular
momentum is neglected, since it is lost by dynamical friction
through mergers of mass clouds $M_c$ on a time scale
$t_{\mathrm{DF}}\approx 0.2\, (\xi/\ln{\xi})\, t_{\mathrm{dyn}}$,
where $\xi=M_H/M_c$ (e.g., Mo \& Mao 2004); major mergers, which are
very frequent at high redshift, imply $\xi \sim$ a few.

The mass of cold gas is increased by cooling of the hot gas
($\dot{M}_{\mathrm{cond}}$), and decreased by star formation
($\dot{M}_{\star}$) and by the energy feedback from SNae
($\dot{M}_{\mathrm{cold}}^{SN}$) and quasar activity
($\dot{M}_{\mathrm{cold}}^{QSO}$). Its evolution thus obeys the
equation:
\begin{equation}\label{eq|mdotcold}
\dot{M}_{\mathrm{cold}} = \dot{M}_{\mathrm{cond}}-
(1-\mathcal{R})\dot{M}_{\star} -
\dot{M}_{\mathrm{cold}}^{SN}-\dot{M}_{\mathrm{cold}}^{QSO}~,\\
\end{equation}
where $\mathcal{R}$ is the fraction of gas restituted to the cold
component by the evolved stars, amounting to $\mathcal{R}\approx
0.3$ under the assumption of instantaneous recycling. Strictly
speaking, this value of $\mathcal{R}$ is an upper limit, since
only a fraction of evolved stars have a significant mass loss in
the evolutionary phases considered here; the results, however, are
only very weakly sensitive to the chosen value, in the allowed
range. The cold mass ending up in the reservoir around the central
super-massive BH is lost at a very low rate (see Granato et al.
2004) and has been neglected in the above equation.

The energy feedback from SNae is parameterized as
\begin{equation}\label{eq|mdotsn}
\dot{M}_{\mathrm{cold}}^{SN} = \beta_{SN}\, \dot{M}_{\star}~,
\end{equation}
where the efficiency of gas removal
\begin{equation}\label{eq|betasn}
\beta_{SN}=\frac {N_{SN}\,
\epsilon_{SN}\,E_{SN}}{E_{\mathrm{bind}}}\approx 0.35
\left(\frac{N_{SN}}{8\times 10^{-3}/ M_{\odot}}\right)
\left(\frac{\epsilon_{SN}}{0.05}\right)
\left(\frac{E_{SN}}{10^{51}\mathrm{erg}}\right)
\left(\frac{M_H}{10^{12} M_{\odot}}\right)^{-2/3} \left(
\frac{1+z}{7}\right)^{-1}
\end{equation}
depends on the number of SNae per unit solar mass of condensed stars
$N_{SN}$, on the energy per SN available to remove the cold gas
$\epsilon_{SN}\,E_{SN}$, and on the specific binding energy of the
gas within the DM halo, $E_{\mathrm{bind}}$. Following Zhao et al.
(2003) and Mo \& Mao (2004), the latter quantity has been estimated
for $z\ga 1$ as $E_{\mathrm{bind}}=V_{\mathrm{H}}^{2}\,
f(c)\,(1+f_{\mathrm{cosm}})/2\approx 5.6\times 10^{14}\,
(M_H/10^{12}\, M_{\odot})^{2/3}\, [(1+z)/7]\,
{\mathrm{cm}^2~\mathrm{s}^{-2}}$; here $V_H$ is the halo circular
velocity at the virial radius and $f(c)\approx
2/3+(c/21.5)^{0.7}\sim 1$ is a weak function of the halo
concentration $c$. We adopt the strength of SN feedback
$\epsilon_{SN}=0.05$ used by Lapi et al. (2006) in order to
reproduce LFs of galaxies and quasars at high redshifts, and also to
reproduce the fundamental correlations between local elliptical
galaxies and dormant BHs.

The cold gas turns into stars at a rate
\begin{equation}\label{eq|aSFR}
\dot{M_{\star}}=\int
\frac{\mathrm{d}M_{\mathrm{cold}}}{\max[t_{\mathrm{cool}},t_{\mathrm{dyn}}]}
\approx {M_{\mathrm{cold}}\over t_{\star}}~,
\end{equation}
where now $t_{\mathrm{cool}}$ and $t_{\mathrm{dyn}}$ refer to the
mass shell $\mathrm{d}M_{\mathrm{cold}}$, and $t_{\star}$ is the
star formation timescale averaged over the mass distribution.
Eqs.~(\ref{eq|mdotinf}) and (\ref{eq|mdotcold}) can be easily
solved setting $\dot{M}_{\star}\approx M_{\mathrm{cold}}/t_{\star}$
and neglecting the effects of the energy injected in the gas by the
accretion onto the central BH. With this approximation the infalling
mass declines exponentially as $M_{\mathrm{inf}}(t) =
M_{\mathrm{inf}}(0)\, e^{-t/t_{\mathrm{cond}}}$, while the SFR
evolves according to
\begin{equation}\label{eq|sfr}
\dot{M}_{\mathrm{\star}}(t)=\frac
{M_{\mathrm{inf}}(0)}{t_{\mathrm{cond}}(\gamma-1/s)} \left[
e^{-t/t_{\mathrm{cond}}}- e^{-s\,\gamma\,
t/t_{\mathrm{cond}}}\right]~,
\end{equation}
with $\gamma\equiv 1-\mathcal{R}+\beta_{SN}$. The quantity $s\equiv
t_{\mathrm{cond}}/t_{\mathrm{\star}}$ is the ratio between the
timescale for the large-scale infall estimated at the virial radius,
and the star formation timescale in the central region. An
isothermal density profile yields $s\approx 5$.

Although eq.~(\ref{eq|sfr}) has been obtained neglecting the quasar
feedback, it turns out to be, at the redshifts relevant here ($z\ga
2$), a good approximation of the results of the full numerical
integration of the system of differential equations given by Lapi et
al. (2006), provided that: (i) we take into account that the star
formation is strongly suppressed by the quasar feedback after a time
\begin{equation}\label{eq|dtburst}
\Delta t_{\mathrm{burst}}\approx 2.5\times 10^8 \, \left({1+z\over
7}\right)^{-1.5}\, \mathcal{F}\left({M_H\over 10^{12}\,
M_{\odot}}\right)~~\mathrm{yr}~,
\end{equation}
where $\mathcal{F}(x)=1$ for $x\geq 1$ and
$\mathcal{F}(x)=x^{-1}$ for $x\leq 1$; (ii) we adopt the
following expression for the condensation timescale
\begin{equation}\label{eq|dtcond}
t_{\mathrm{cond}}\approx 4\times 10^8\,
\left(\frac{1+z}{7}\right)^{-1.5}\, \left(\frac {M_H}{10^{12}\,
M_{\odot}}\right)^{0.2}~~\mathrm{yr}~.
\end{equation}
In the latter expression, the scaling with the redshift mirrors the
dependence of the cooling and/or dynamical time on redshift.
Moreover, the weak dependence on $M_H$ renders the impact of the
energy feedback from the quasar on the infalling gas; its effect is
mimicked by allowing longer condensation timescales for the gas in
more massive halos, exposed to stronger quasar feedback from more
massive BHs.

Note that the SFR evolves differently for different halo masses,
see Fig.~\ref{fig|sfr:mstar}.
For large masses $M_H\ga 10^{12}\, M_{\odot}$,
it increases almost linearly with galaxy age in the initial stages
$t\ll t_{\rm cond}$, and then it is suddendly stopped by the energy feedback from the quasar
at $t\approx \Delta t_{\rm burst}$. On the contrary, for low-mass halos
$M_H\la 10^{12}\, M_{\odot}$ it is first almost constant and then slowly
declining; indeed, in such low mass galaxies the quasar feedback is very mild
because the mutual effects of star formation and gas flow
toward the reservoir around the central BH result in a strong
decline of the BH mass with decreasing halo mass [e.g., Shankar et
al. 2006, their eq.~(16)]; thus the SFR can proceed for much longer times
(note from eq.~(A7) that $\Delta t_{\rm burst} \ga 3$ Gyr at $z\approx 6$
for $M_H\la 10^{11}\, M_{\odot}$).
Note that Fig.~\ref{fig|sfr:mstar} refers to redshift $z=6$;
at $z\approx 3$ the SFR retains the same time dependence as at $z=6$,
but is on average lower by a factor of $\approx 2$ for a given halo mass.

The total mass $M_{\star}^{\mathrm{burst}}$ cycled through stars
during the time $\Delta t_{\mathrm{burst}}$ can be estimated by
integrating the SFR [eq.~(\ref{eq|sfr})]. The present-day total mass
of surviving stars $M_{\star}^{\mathrm{now}}\equiv
f_{\mathrm{surv}}\, M_{\star}^{\mathrm{burst}}$ can then be derived
by assuming a specific IMF; $f_{\mathrm{surv}}$ is around $60\%$
after $10$ Gyr from the burst for a Salpeter IMF, and is around
$30\%$ for the IMF adopted here. The latter is a double power-law
with slope $1.25$ from $120\, M_{\odot}$ to $1\, M_{\odot}$ and
$0.4$ from $1\, M_{\odot}$ down to $0.1\, M_{\odot}$ (Romano et al.
2002; for a review on the IMF see Chabrier 2005).

An interesting outcome of our model is that the fraction of the
original baryons associated to the galactic halo that is trapped in
present-day stars stays almost constant at
$M_{\star}^{\mathrm{now}}/M_{\mathrm{inf}}(0) \approx 0.2$ for $M_H
\ga 3 \times 10^{11}\, M_{\odot}$, and decreases very rapidly below
this threshold. As discussed by Shankar et al. (2006), this is in
keeping with results obtained through X-ray and weak lensing
estimates of halos around galaxies.

When computing the metal content of the cold gas, one has to take
into account the enrichment of the primordial infalling gas due to
earlier generations of stars, and the gas outflows due to the energy
injection by SNae and quasars. In the first stages, the chemical
enrichment of the cold gas component is very rapid; e.g., $1/100$
and $1/10$ the solar abundance in the gaseous component
is attained after $\sim 1.2\times 10^7$ and $\sim 4\times 10^7$ yr,
almost independently of the halo mass and redshift.  Only less
than $1\%$ and $10\%$ of the total stellar mass has been built up at
these times and therefore the expected final mass fraction of
metal-poor stars is tiny.
This rapid enrichment is due to the short lifetime of massive stars
$\ga 10\, M_{\odot}$ (that are efficient metal producers) relative to the
timescale $t_{\rm cond}$ of the infall of the diffuse medium with
primordial composition, which dilutes the cold star-forming gas.
A simple approximation of the subsequent evolution reads
\begin{equation}\label{eq|metallicity}
Z(t)\approx 10^{-2}\, {t\over 7\times 10^7~ \mathrm{yr}}\,
\left({1+z\over 7}\right)^{5/4}, ~~~~ \mathrm{for}~~ 7\times
10^7\, \mathrm{yr}\la t\la \min[\Delta t_{\mathrm{burst}},\Delta
t_{\mathrm{sat}}]~,
\end{equation}
where $\Delta t_{\mathrm{sat}}=7\times 10^7 \, ({M_H/10^{10}\,
M_{\odot}})^{2/3}$ yr is a characteristic time after which $Z$ keeps
almost constant.

\section{Reionization}

In this Appendix we briefly describe the computation of the
reionization redshift predicted by our model.  The average global
production rate of ionizing photons per unit volume at the cosmic
time $T$ is
\begin{equation}\label{eq|ndotioncosm}
\langle\dot{N}_{912}\rangle (T) =
\int{\mathrm{d}\dot{N}_{912}}\,\dot{N}_{912}\,\Phi(\dot{N}_{912},T)~,
\end{equation}
where $\dot{N}_{912}$ is given by eq.~(\ref{eq|ndotion}) and its
distribution function $\Phi(\dot{N}_{912},T)$ can be derived in
strict analogy to eq.~(\ref{eq|UVcounts}). The transition from a
neutral to a reionized IGM can be described in terms of the
evolution of the volume filling factor of HII regions,
$Q_{\mathrm{HII}}$, which is ruled by the equation (Shapiro \&
Giroux 1987; Madau, Haardt \& Rees 1999)
\begin{equation}\label{eq|filling}
\frac{\mathrm{d}Q_{\mathrm{HII}}}{\mathrm{d}T}=
\frac{\langle\dot{N}_{912}\rangle}{n_{\mathrm{H}}}-\frac{Q_{\mathrm{HII}}}{t_{
\mathrm{rec}}};
\end{equation}
here $n_{\mathrm{H}}$ is the comoving $\mathrm{HI}$ number
density, and the recombination time is given by (Madau, Haardt \&
Rees 1999)
\begin{equation}\label{eq|trec}
t_{\mathrm{rec}}=0.3\,\left(\frac{1+z}{4}\right)^{-3}\,
\left(\frac{\mathcal{C}}{10}\right)^{-1}~ \mathrm{Gyr},
\end{equation}
in terms of the clumping factor of the IGM, $\mathcal{C}$. The IGM
clumping factor has been investigated through hydrodynamical
simulations (e.g., Gnedin \& Ostriker 1997). Iliev et al. (2006)
proposed an analytic description of the simulation results:
$\mathcal{C}(z)=17.6\,e^{-0.1\,z+0.0011\,z^2}$. In the relevant
redshift interval $7\la z\la 10$ one finds $10\ga \mathcal{C} \ga
7$. Here we adopt $\mathcal{C}=7$ as a reference value. Reionization
is completed at the redshift when $Q_{\mathrm{HII}}=1$ (right
panel of Fig. \ref{fig|reion_extgigi}).
Finally, the optical depth to electron scattering is given by
\begin{equation}\label{eq|optdepth}
\tau_{\mathrm{es}}(z)=\int_0^z{\mathrm{d}z'}~\left|\frac{\mathrm{d}T}{\mathrm{d}
z'}\right|\, c\,\sigma_{\mathrm{T}}\, n_e(z'),\label{eq|eloptdep}
\end{equation}
where $n_e=Q_{\mathrm{HII}}\, n_B\, (1+z)^3$ is the electron
density, $n_B\approx 2.5\times 10^{-7}$ cm$^{-3}$ is the present day
baryon density, and $\sigma_{\mathrm{T}}\approx 6.65\times 10^{-25}$
cm$^2$ is the Thomson cross section.

\end{appendix}

\end{document}